\documentclass[aps,superscriptaddress]{revtex4}
\usepackage[colorlinks=true, pdfstartview=FitV, linkcolor=blue, citecolor=red, urlcolor=magenta, breaklinks=true, pdftex]{hyperref}
\usepackage{graphicx}  
\usepackage{amsfonts}
\begin{document}
\title{Quantum-corrected rotating acoustic black holes in Lorentz-violating background}
\author{M. A. Anacleto}
\email{anacleto@df.ufcg.edu.br}
\affiliation{Departamento de F\'{\i}sica, Universidade Federal de Campina Grande, Caixa Postal 10071, 58429-900 Campina Grande, Para\'{\i}ba, Brazil}
\author{F. A. Brito}
\email{fabrito@df.ufcg.edu.br}
\affiliation{Departamento de F\'{\i}sica, Universidade Federal de Campina Grande, Caixa Postal 10071, 58429-900 Campina Grande, Para\'{\i}ba, Brazil}
\affiliation{Departamento de F\'isica, Universidade Federal da Para\'iba, Caixa Postal 5008, 58051-970 Jo\~ao Pessoa, Para\'iba, Brazil}
\author{C. V. Garcia}
\email{carol.v.garcia@hotmail.com}
\affiliation{Departamento de F\'{\i}sica, Universidade Federal de Campina Grande, Caixa Postal 10071, 58429-900 Campina Grande, Para\'{\i}ba, Brazil}
\author{G. C. Luna}
\email{gabrielacluna@hotmail.com}
\affiliation{Departamento de F\'isica, Universidade Federal da Para\'iba, Caixa Postal 5008, 58051-970 Jo\~ao Pessoa, Para\'iba, Brazil}
\author{E. Passos}
\email{passos@df.ufcg.edu.br}
\affiliation{Departamento de F\'{\i}sica, Universidade Federal de Campina Grande, Caixa Postal 10071, 58429-900 Campina Grande, Para\'{\i}ba, Brazil}
%\email{anacleto, fabrito, passos@df.ufcg.edu.br, gabrielacluna@hotmail.com}

\begin{abstract}
In this paper we explore the effect of the generalized uncertainty principle and modified dispersion relation to compute Hawking radiation from a rotating acoustic black hole in the tunneling formalism by using the Wentzel-Kramers-Brillouin (WKB) approximation applied to the Hamilton-Jacobi method. The starting point is to consider the planar acoustic black hole metric found in a Lorentz-violating Abelian Higgs model. In our analyzes we investigate quantum corrections for the Hawking temperature and entropy. A logarithmic correction and an extra term that depends on a conserved charge were obtained. We also have found that the
changing in the Hawking temperature ${\cal T}_H$ for a dispersive medium due to a Lorentz-violating background accounts for supersonic
velocities in the general form $(v_g-v_p)/v_p = \Delta {\cal T}_H/{\cal T}_H\sim10^{-5}$ in Bose-Einstein-Condensate (BEC) systems. 

\end{abstract}
\maketitle
\pretolerance10000

\section{Introduction}
In modern physics the attempt to construct a consistent theory of quantum gravity arising when combined general relativity and quantum mechanics has been extensively explored in the literature. This theory would be important for a better understanding of the final stage of a black hole.
Several theories have been put forward for this purpose among which we highlight the loop quantum gravity and string theory which include expected features for a consistent theory.
These theories present some points in common such as the existence of a minimum length~\cite{Garay:1994en,AmelinoCamelia:2000ge,Casadio:2014pia}, as a consequence we have the modification of the principle of uncertainty of Heisenberg, the so-called generalized Heisenberg uncertainty principle (GUP)~\cite{ADV, Tawfik:2014zca, KMM, Tawfik:2015kga, Gangopadhyay:2015zma, GDS2014}.
The interest in the study of Hawking radiation in models considering the GUP has increased 
a lot in the last years~\cite{Brustein,Kim2006,Park2007,Sun2004,Yoon2007,XLi,zhao2004,RZ2003,WK2006,KKP2007,JHa2007,YWKim2007,Anacleto:2015mma,
Anacleto:2015kca,ABBS2015,Chen:2017kpf,Ovgun:2015box,Gecim:2018sji,Li:2016mwq,Sadeghi:2016xym,
Casadio:2017sze,Maluf:2018lyu,Silva:2017gki,Gecim:2017zid,Nozari:2008gp,Nozari:2012nf,Santos:2015gja}.
In particular the analysis of Hawking radiation in analog models has been explored extensively in the literature mainly due to the possibility of  being tested in laboratory~\cite{Robertson:2012ku}.

The fact of the Hawking radiation depending only on the kinematic properties of the space-time background led Unruh in 1981, to propose a theoretical method of analogous gravity~\cite{Unruh} with the aim of producing kinematic conditions similar to that of a black hole.
Since then, several works in analog models have been explored ~\cite{RS, Mathis, UL, Philbin, RSch, Novello, Garay, OL, SG}.
{One such application has been the study of quasi-normal modes in which it has been shown recently~\cite{Patrick:2018orp} that by considering an extra structure (core structure of the vortex) in the study of rotating acoustic black hole, the spectrum of quasi-normal modes is changed, and also a possible experimental setup was suggested by the authors in order to test the results obtained. }
In addition, starting from relativistic models the metric of acoustic black holes has been 
determined~\cite{Xian, Bilic, Liberati, Molina, ABP12, ABP11, ABP10, Anacleto:2013esa}.
In~\cite{ABP,Anacleto:2012du,Anacleto:2015mta,Anacleto:2016ukc,Anacleto:2018acl} these effective metrics were applied to the study of the phenomena of superresonance, absorption and the analogous Aharonov-Bohm effect.

In studies proposed by Steinhauer~\cite{Steinhauer:2014dra, Steinhauer:2015ava} through the use of analog models one has shown great progress in the possibility of detecting the Hawking radiation in laboratory.
An effective way of determining Hawking radiation is to apply the Hamilton-Jacobi method which is based on the tunneling process of elementary particles across the black hole event horizon~\cite{Parikh,Vagenas:2002hs}.
In this method the WKB approximation is applied in order to determine the imaginary contribution of the action.
In~\cite{Becar:2010zza} by applying this approach it was investigated the Hawking radiation of an acoustic black hole.
Also, the authors in~\cite{Sakalli:2016mnk, Anacleto:2015awa} analyzed the Hawking radiation of a rotating acoustic black hole in the tunneling formalism with GUP. In~\cite{Zhao, Anacleto:2014apa, Anacleto:2016qll} by investigating the Hawking radiation of a rotating acoustic black hole, the authors also explored the effect of the GUP by applying the brick wall method.

The purpose of the present work is to apply the Hamilton-Jacobi method and the WKB approximation within the tunneling formalism to examine the Hawking radiation and the entropy of a rotating acoustic black hole metric obtained from an Abelian Higgs model in a Lorentz-violating background.
%For this purpose we will consider the GUP and also the modified dispersion relation derived from the model itself to calculate the Hawking temperature and the corrected entropy.
{For this purpose we will consider two approaches to calculate quantum corrections for the Hawking temperature and entropy. First we will apply the GUP in order to determine a modified dispersion relation. In the second case we will focus on a dispersion relation that is modified but the uncertainty principle preserves the standard Heisenberg form.}

Unlike Hawking radiation, which is a purely kinematic effect, an analogous form for the Bekenstein-Hawking entropy has been little known until recently. However, in \cite{Rinaldi} has been argued that such an analogy arises in a Bose-Einstein condensate system such that the Bekenstein-Hawking entropy can be understood as an {\it entanglement entropy}. 
{In a Bose-Einstein condensate system entanglement entropy is related to the phonons that are created by the Hawking mechanism. 
In this case the entropy of the acoustic black hole depends on the area of the event horizon.
In addition, this issue has been addressed by J. Steinhauer in~\cite{Steinhauer:2015ava}.
He has observed the entanglement of Hawking radiation from an acoustic black hole due to quantum vacuum fluctuations in a Bose-Einstein condensate. 
It has also been found that there is a correlation between the particles outside and the partner particles within the black hole. 
Such entanglement is reduced as energy decreases. 
Also in~\cite{Giovanazzi:2011az} the entanglement entropy of an acoustic black hole was investigated in 1D degenerate ideal fermi fluids,  which entropy flows from the sonic horizon to subsonic and supersonic regions. These examples in ultra cold quantum gases certainly lead us to look for other similar systems that can present entropy entanglement. Despite the difficulty of finding Hawking radiation from astrophysical black holes, such radiation can be found experimentally in analogue black holes \cite{Steinhauer:2014dra}.  Thus, for the same reason, it is interesting to propose similar investigations of analogue Bekenstein-Hawking entropy in laboratory.
 }
For a comprehensive study on entaglement entropy and black holes see \cite{Solodukhin:2011gn} --- see also \cite{Anacleto:2016qll} for a more recent discussion on related issues.

The paper is organized as follows. In Sec.~\ref{II} we make a short revision on the computation of the acoustic metric from the Abelian Higgs model. In Sec.~\ref{HJ-method} we apply the Hamilton-Jacobi method to calculate the Hawking temperature for a rotating acoustic black hole. In Sec.~\ref{stat-ent} we consider the generalized uncertainty principle (GUP) and modified dispersion relation to compute quantum corrections to Hawking temperature and entropy. Finally in Sec.~\ref{conclu} we present our final considerations.

\section{The Lorentz-Violating Acoustic Black Hole}
\label{II}
{In this section, we shall focus on the planar acoustic black hole metrics to investigate the corrections of the Hawking temperature and the entropy.
The starting point is to consider the following Lagrangian of the Lorentz-violating Abelian Higgs model 
\begin{eqnarray}
\label{acao}
{\cal L}&=&-\frac{1}{4}F_{\mu\nu}F^{\mu\nu} +|D_{\mu}\phi|^2+ m^2|\phi|^2-b|\phi|^4+ k^{\mu\nu}D_{\mu}\phi^{\ast}D_{\nu}\phi,
\end{eqnarray}
where $F_{\mu\nu}=\partial_{\mu}A_{\nu}-\partial_{\nu}A_{\mu}$, $D_{\mu}\phi=\partial_{\mu}\phi - ieA_{\mu}\phi$ and $k^{\mu\nu}$  is a constant tensor introducing the Lorentz symmetry breaking  given by
\begin{equation}
k_{\mu\nu}=\left[\begin{array}{clcl}
\beta &\alpha &\alpha & \alpha\\
\alpha &\beta &\alpha &\alpha \\
\alpha &\alpha &\beta &\alpha\\
\alpha &\alpha &\alpha &\beta
\end{array}\right], \quad(\mu,\nu=0,1,2,3),
\end{equation}
with $\alpha$ and $\beta$ being real parameters. 

Next we will briefly show some steps to obtain the acoustic metric from the Lagrangian (\ref{acao}).
The first step is to represent the scalar field as $\phi = \sqrt{\rho(x, t)} \exp {(iS(x, t))}$, and so the Lagrangian is put into the form
\begin{eqnarray}
&&{\cal L} = -\frac14 F_{\mu\nu}F^{\mu\nu} + \rho\partial_{\mu}S\partial^{\mu}S - 2e\rho A_\mu \partial^{\mu}S+ e^2\rho A_\mu A^\mu + m^2\rho - b\rho^2\nonumber\\
&+& k^{\mu\nu} \rho(\partial_{\mu}S\partial_{\nu}S-2eA_\mu\partial_{\nu}S+eA_\mu A_\nu)
+\frac{\rho}{\sqrt{\rho}}(\partial_{\mu}\partial^{\mu}+k^{\mu\nu}\partial_{\mu}\partial_{\nu})\sqrt{\rho}.
\end{eqnarray}
In the second step we linearize the equations of motion around the background $(\rho_0,S_0)$, with $\rho=\rho_0+\rho_1$ and  $S=S_0+\psi$ and finally we obtain the equation of motion for a linear acoustic disturbance $\psi$ given by a Klein-Gordon equation in a curved space
\begin{eqnarray}
\frac{1}{\sqrt{-g}}\partial_{\mu}(\sqrt{-g}g^{\mu\nu}\partial_{\nu})\psi=0,
\end{eqnarray}
where $g_{\mu\nu}$ is precisely the effective acoustic metric. In this work we will restrict our analyzes only to the case where $ \beta=0 $ 
and $ \alpha\neq 0 $. The case where $ \beta\neq 0 $ and $ \alpha=0$ has been investigated in~\cite{Sakalli:2016mnk}.

In this case the line element of the acoustic metric in 2 + 1 dimensions in the non-relativistic limit, and keeping terms up to first order in $ \alpha $, is given by \cite{ABP11,ABP10}
\begin{eqnarray}
\label{ECC}
ds^2=-\tilde{\alpha}\left(c^2_s-\frac{v^2}{\tilde{\alpha}}\right)d t^2 -2\big(\vec{v}\cdot d\vec{x}\big) dt
+\big[1+2\alpha \big(v_x+v_y\big)\big]d\vec{x}^2,
\end{eqnarray}
where $ c_s=\sqrt{dh/d\rho} $ is the sound velocity in the fluid and $ v $ is the fluid velocity.
The metric above can be rewritten in polar coordinates as follows
\begin{eqnarray}
\label{ECP}
ds^2=-\tilde{\alpha}\left(1-\frac{v^2}{\tilde{\alpha}c^2_s}\right)d t^2+\tilde{\alpha}^{-1}\left(1-\frac{v^2_r}{\tilde{\alpha}c^2_s}\right)^{-1}d r^2 -\frac{2 v_{\phi}}{c_s} rd\phi dt 
+\big[1+2\alpha \big(v_{r}+v_{\phi}\big)\big]dr^2d\phi^2.
\end{eqnarray}
In polar coordinates the velocity profile of the fluid is 
\begin{eqnarray}\label{vab}
\vec{v}=\frac{A}{r}\hat{r}+\frac{B}{r}\hat{\phi},
\end{eqnarray} 
where $ A $ and $ B $ are constants related to circulation and draining respectively. Thus, the metric of the rotating acoustic black hole takes the following form \cite{ABP11}
\begin{eqnarray}
\label{ELB}
ds^2=-\tilde{\alpha}\left(1-\frac{\tilde{r}_{e}^2}{r^2}\right)d t^2
+\tilde{\alpha}^{-1}\left(1-\frac{\tilde{r}_{h}^2}{r^2}\right)^{-1}dr^2
-\frac{2B}{c_s}d\phi d t+\left[1+\frac{2\alpha(\tilde{\alpha}^{1/2}c_sr_{h}+B)}{r}\right]r^2d\phi^2,
\end{eqnarray}
where $\tilde\alpha=1+\alpha$, $ \tilde{r}_e $ is the radius of the ergosphere given by,
 \begin{eqnarray}
\tilde{r}_{e}=\frac{r_e}{\sqrt{\tilde{\alpha}}}, \quad r_e=\sqrt{{r}_{h}^2+\frac{B^2}{c^2_s}}, 
\end{eqnarray}
and $ \tilde{r}_h $ is the horizon, that is
\begin{eqnarray}
\tilde{r}_h=\frac{r_h}{\sqrt{\tilde{\alpha}}}, \quad r_{h}=\frac{|A|}{c_s}.
\end{eqnarray}
}
\section{Hamilton-Jacobi Method and WKB approximation}
\label{HJ-method}

{In this section we consider the WKB approximation in the tunneling formalism and apply the Hamilton-Jacobi method to calculate the Hawking temperature for an acoustic black hole.
}
\subsection{Acoustic metric with $ B= 0 $}

{We now consider the case $B=0$ (no rotation), so we can display the metric in stationary form as follows
\begin{eqnarray}
\label{ELBNR}
ds^2=-f(r)d t^2+\frac{1}{f(r)}dr^2+\left(1+\frac{\eta}{r} \right) r^2d\phi^2,
\end{eqnarray}
where $ f(r)= \tilde{\alpha}\left(1-\frac{\tilde{r}^2_h}{r^2} \right)$ and $\eta=2\alpha(\tilde{\alpha}^{1/2}\tilde{r}_{h}) $, with $ c_s=1 $. 
Thus, we can determine the Hawking temperature in terms of $\tilde{r}_h $ as follows
\begin{eqnarray}
\tilde{T}_{h}=\frac{f^{\prime}(\tilde{r}_h)}{4\pi}=\frac{\tilde{\alpha}}{2\pi \tilde{r}_{h}}=\frac{\left(1+\alpha\right)}{2\pi \tilde{r}_{h}}.
\end{eqnarray}
We can also express the result in terms of $ r_h $
\begin{eqnarray}\label{Htemp}
\tilde{T}_{h}=\tilde{\alpha}^{3/2} T_h, 
\end{eqnarray}
where $T_h=(2\pi {r}_{h})^{-1}$ is the Hawking temperature of the acoustic black hole for $\alpha=0$. 

In order to find the Hawking temperature by the Hamilton-Jacobi method, we start with the Klein-Gordon equation for a scalar field $ \Phi $ in the curved space given by 
\begin{eqnarray}
\left[\frac{1}{\sqrt{-g}}\partial_{\mu}\left(\sqrt{-g}g^{\mu\nu}\partial_{\nu}\right)-\frac{m^2}{\hbar^2}\right]\Phi=0,
\end{eqnarray}
where $m$ is the mass of a scalar particle. Then using the WKB approximation
\begin{eqnarray}
\Phi=\exp \left[\frac{i}{\hbar}{\cal I}(t,r,\phi)\right],
\end{eqnarray}
we find
\begin{eqnarray}
\label{wkb}
g^{\mu\nu}\partial_{\mu}{\cal I}\partial_{\nu}{\cal I}+m^2=0.
\end{eqnarray}
Hence, we can rewrite equation (\ref{wkb}) in the metric (\ref{ELBNR}) in the form
\begin{eqnarray}
-\frac{1}{f(r)}(\partial_{t}{\cal I})^2+f(r)(\partial_{r}{\cal I})^2+\frac{1}{(1+\eta/r) r^2}(\partial_{\phi}{\cal I})^2+m^2=0.
\end{eqnarray}
Taking into account the symmetry of the metric we will assume a solution to the equation above that reads 
\begin{eqnarray}
{\cal I}=-Et  + W(r) + J_{\phi}\phi,
\end{eqnarray}
where
\begin{eqnarray}
\partial_t {\cal I}=-E, \quad \partial_r {\cal I}=\frac{dW(r)}{dr}, \quad \partial_{\phi} {\cal I}=J_{\phi},
\end{eqnarray}
and $J_{\phi}$ is a constant. Thus, for the classical action we have 
\begin{eqnarray}
{\cal I}=-Et  + \int dr\frac{\sqrt{E^2-f(r)(m^2-\frac{J^2}{(1+\eta/r) r^2}})}{f(r)}  + J_{\phi}\phi.
\end{eqnarray}
In the near horizon regime, $ r\rightarrow \tilde{r}_h $, we will apply the following approximation
$ f(r)\approx 2\kappa(r-\tilde{r}_h)$, where $\kappa=f^{\prime}(\tilde{r}_h)/2 $ is the surface gravity of acoustic black hole. In this case the spatial part of the action function, becomes
\begin{eqnarray}
W(r)=\frac{1}{2\kappa}\int dr\frac{\sqrt{E^2-2\kappa(r-\tilde{r}_h)(m^2-\frac{J^2}{\delta r^2}})}{(r-\tilde{r}_h)}
=\frac{2{\pi} i E}{2\kappa}.
\end{eqnarray}
Therefore, we can calculate the probability of tunneling by applying the following equation 
\begin{eqnarray}
\Gamma\simeq \exp[-2Im {\cal I}]=\exp\left[-\frac{2{\pi}  E}{\kappa}\right].
\end{eqnarray}
On the other hand,  comparing the above result with the Boltzmann factor $ \Gamma\simeq\exp (-E/\tilde{T}_h) $, we determine the Hawking temperature of the acoustic black hole
\begin{eqnarray}
\label{th}
\tilde{T}_h=\frac{\kappa}{2\pi}=\frac{\tilde{\alpha}}{2\pi \tilde{r}_h}.
\end{eqnarray}
Moreover, equation (\ref{th}) can be expressed in terms of $ r_h $ as being 
\begin{eqnarray}
\tilde{T}_h=\tilde{\alpha}^{3/2}T_h.
\end{eqnarray}
Note that this result for the temperature coincides with that obtained earlier in (\ref{Htemp}).
}

\subsection{Acoustic metric with $ B\neq 0 $}

Let us now consider the case with rotation ($ B\neq 0 $). Thus the metric (\ref{ELB}) can be written as follows
\begin{eqnarray}
ds^2=-\left[f(r) -\frac{\gamma B^2}{r^3}\right]d t^2+\frac{1}{f(r)}dr^2+\left(1+\frac{\gamma}{r} \right) r^2d\varphi^2,
\end{eqnarray}
where, $ \gamma= 2\alpha(\tilde{\alpha}^{1/2}\tilde{r}_{h}+B)$ and 
\begin{equation}
d\varphi=d\phi - \frac{B dt}{(1-\gamma/r)r^2},
\end{equation}
is the coordinate transformation used to leave the metric in the diagonal form.
{Then, following the same steps as described above to find the Hawking temperature, the probability of tunneling near the event horizon is 
\begin{eqnarray}
\Gamma\simeq \exp\left[-{4{\pi}  E}/{\tilde{\kappa}}\right],
\end{eqnarray}
where 
\begin{eqnarray}
 \tilde{\kappa} =\sqrt{\left[f^{\prime}(\tilde{r}_h)+3\gamma B^2/\tilde{r}_h^4 \right]f^{\prime}(\tilde{r}_h)},
\end{eqnarray}
is the surface gravity and finally comparing $ \Gamma $ with the Boltzmann factor
$ \exp(-E/{\cal T}_h) $, we have the Hawking temperature given by
\begin{eqnarray}
{\cal T}_h&=&\frac{\tilde{\kappa}}{4\pi}=\frac{\sqrt{\left[f^{\prime}(\tilde{r}_h)+3\gamma B^2/\tilde{r}_h^4 \right]f^{\prime}(\tilde{r}_h)}}{4\pi}
=\frac{\tilde{\alpha}}{2\pi \tilde{r}_h}\sqrt{1+\frac{3\gamma B^2}{2\tilde{\alpha} \tilde{r}_h^3}},
\nonumber\\
&=&\frac{\tilde{\alpha}}{2\pi \tilde{r}_h} +\frac{3\gamma B^2}{8\pi\tilde{r}^4_h}+{\cal O}(\alpha^2)
=\frac{\tilde{\alpha}}{2\pi \tilde{r}_h}+\frac{3\alpha\sqrt{\tilde{\alpha}} B^2}{4\pi\tilde{r}^3_h} 
+\frac{3\alpha B^3}{4\pi\tilde{r}^4_h}+{\cal O}(\alpha^2),
\end{eqnarray}
where for $ B=0 $ we recover the result obtained in (\ref{th}).}
We can also write the temperature in terms of area $\tilde{A}=2\pi\tilde{r}_h$ as
\begin{eqnarray}
\label{thbgup}
{\cal T}_h&=&\frac{\tilde{\alpha}}{\tilde{A}}+\frac{6\pi^2\alpha\sqrt{\tilde{\alpha}} B^2}{\tilde{A}^3} 
+\frac{12\pi^3\alpha B^3}{\tilde{A}^4}+{\cal O}(\alpha^2).
\end{eqnarray}

\section{Quantum-corrections to the entropy}
\label{stat-ent}
In this section we consider the generalized uncertainty principle (GUP)~\cite{ADV, Tawfik:2014zca, KMM, Tawfik:2015kga, Gangopadhyay:2015zma}, 
which is defined as
\begin{eqnarray}
\label{gup}
\Delta x\Delta p\geq \hbar\left( 1-\frac{\lambda l_p}{\hbar} \Delta p +\frac{\lambda^2 l^2_p}{\hbar^2} (\Delta p)^2 \right),
\end{eqnarray}
where $\lambda$ is a dimensionless positive parameter and $ l_p $ is the Planck length.

Now the equation (\ref{gup}) can be recast in the form
\begin{eqnarray}
\Delta p\geq \frac{\hbar(\Delta x +\lambda l_p)}{2\lambda^2 l_p^2}
\left(1- \sqrt{1-\frac{4\lambda^2 l_p^2}{(\Delta x +\lambda l_p)^2}}\right).
\end{eqnarray}
%Taking into account that $ l_p\ll 1$ and so without loss of generality 
In the following computations, without loss of generality, we shall adopt the units $ G=c=k_B=\hbar=l_p=1 $.  Now we perform a power series in $\lambda$ to obtain
\begin{eqnarray}
\label{p}
\Delta p\geq \frac{1}{2\Delta x}\left[1-\frac{\lambda}{2\Delta x}+ \frac{\lambda^2}{2(\Delta x)^2}+\cdots    \right].
\end{eqnarray}
For the case without GUP, that is, when $ \lambda=0 $ we have the Heisenberg uncertainty principle
\begin{eqnarray}
\Delta x\Delta p\geq 1 .
\end{eqnarray}
From the above equation we can obtain a bound for massless particles given by the relation
\begin{eqnarray}
E\Delta x\geq1.
\end{eqnarray}
In this case, the equation (\ref{p}) can be written as follows
\begin{eqnarray}
\label{rdgup}
{\cal E}\geq E\left[1-\frac{\lambda}{2(\Delta x)}+ \frac{\lambda^2}{2(\Delta x)^2}+\cdots    \right].
\end{eqnarray} 
{In order to obtain a dispersion relation in terms of the momentum we start from the GUP  (\ref{gup}) considering only the quadratic contribution (which is dominant in the case of large moment).  
In this case we find the following dispersion relation
\begin{eqnarray}
\label{rdgup2}
{\cal E}\geq E\left[1+ \frac{\lambda^2}{2(\Delta x)^2}+\cdots    \right].
\end{eqnarray} 
In this way the relation (\ref{rdgup2}) can still be written in terms of the moment assuming that  $ p\sim \Delta p \geq 1/\Delta x $ 
and thus we obtain the following dispersion relation for the {\it supersonic} case}
 \begin{eqnarray}
\label{dispv}
{\cal E}= E\left(1+ \frac{\lambda^2 p^2}{2}\right), \quad \quad {\cal E}^2=E^2\left(1+ \lambda^2 p^2\right).
\end{eqnarray} 
It is important to note that the second term corresponds in our case to a dispersion due to viscosity of the fluid, so the contribution of the GUP introduces a viscosity effect. Further discussions on this subject can be found in~\cite{Richartz:2012bd,Visser:2001jd} -- see also~\cite{Robertson:2012ku}.
{It should be mentioned here that a `superluminal' dispersion relation of type (\ref{dispv}) is obeyed in excitations of a Bose-Einstein condensate~\cite{Robertson:2012ku,LKO}.}
On the other hand starting from a Lagrangian density with terms of high derivatives  we can obtain the following dispersion relation~\cite{Corley:1996ar,Robertson:2012ku,Richartz:2012bd,Anacleto:2013esa}
\begin{eqnarray}
\label{dispdh}
{\cal E}^2 = \left(1\pm \frac{k^2}{\Lambda^2}\right)E^2,
\end{eqnarray} 
where the upper signal corresponds to the superluminal (supersonic for an acoustic black hole) case, the lower signal for the 
subluminal (subsonic for an acoustic black hole) case and $ \Lambda $ is a dispersive momentum scale. 
Hence, at the limit of $ \Lambda\rightarrow \infty $ we obtain $ {\cal E}^2=E^2=k^2 $ which is the dispersion relation for a non-dispersive medium. In \cite{Corley:1996ar} the authors Corley and Jacobson have examined examples of dispersion relation with different types of behavior. From equation (\ref{dispv}) we can obtain the phase ($ v_p $)  and group ($ v_g $) velocities given respectively by
\begin{eqnarray}
v_p=\frac{{\cal E}}{p}=1+ \frac{\lambda^2 p^2}{2},
\end{eqnarray}
and
\begin{eqnarray}
v_g=\frac{d{\cal E}}{dp}=1+ \frac{3\lambda^2 p^2}{2}.
\end{eqnarray}
Applying the Rayleigh's formula that relates the phase and group velocities
\begin{eqnarray}
\label{rf}
v_g=v_p+p\left(\frac{d v_p}{dp}\right),
\end{eqnarray}
gives
\begin{eqnarray}
\frac{v_g-v_p}{v_p}=\lambda^2 p^2.
\end{eqnarray}
This result shows that $ v_g > v_p$, which corresponds to the supersonic case.

\subsection{Acoustic black hole ($ B= 0 $)}

{As in the previous section, the probability of tunneling for a particle with energy ${\cal E}$ is determined by means of the following relation,
\begin{eqnarray}
\Gamma\simeq \exp[-2Im {\cal I}]=\exp\left[\frac{2{\pi} {\cal E}}{a}\right].
\end{eqnarray}
Next we compare with the Boltzmann factor to find the corrected Hawking temperature
\begin{eqnarray}
T=\tilde{T}_h\left[ 1-\frac{\lambda}{2(\Delta x)}+ \frac{\lambda^2}{2(\Delta x)^2}+\cdots   \right]^{-1}.
\end{eqnarray}
Since the minimum uncertainty in our model is of the order of the radius of the horizon so from the equation above the Hawking temperature corrected by the GUP reads
\begin{eqnarray}
T=\frac{\tilde{\alpha}}{2\pi \tilde{r}_h }\left[ 1-\frac{\lambda}{4\tilde{r}_h }+ \frac{\lambda^2}{8\tilde{r}_h^2 }+\cdots   \right]^{-1}.
\end{eqnarray}
In order to obtain the entropy of the acoustic black hole corrected by the GUP we apply the first law of thermodynamics, so we have~\cite{Anacleto:2015awa}
\begin{eqnarray}
\label{entA}
S&=&\int\frac{dE}{T}=\int\frac{\kappa dA}{8\pi T}=\int\frac{d\tilde{A}}{8\pi \tilde{r}_hT}
=\tilde{\alpha}^{-1}\int\frac{d\tilde{A}}{4}\left[ 1-\frac{\pi\lambda}{2\tilde{A}}+ \frac{\pi^2\lambda^2}{2\tilde{A}^2}+\cdots   \right],
\nonumber\\
&=&\tilde{\alpha}^{-1}\left[\frac{\tilde{A}}{4}-\frac{\pi\lambda}{8}\ln{\frac{\tilde{A}}{4}}-\frac{\pi^2\lambda^2}{32 \tilde{A}/4}+\cdots\right].
\end{eqnarray}
This result for the entropy can also be presented in terms of the horizon radius as }
\begin{eqnarray}
\label{entropy-b}
S=\tilde{\alpha}^{-1}\left[\frac{2\pi \tilde{r}_h}{4}-\frac{\pi\lambda}{8}\ln{\frac{2\pi \tilde{r}_h}{4}}
-\frac{\pi^2\lambda^2}{8}\tilde{T}_h+\cdots\right],
\end{eqnarray}
here $ \tilde{A}=2\pi \tilde{r}_h=A/\sqrt{\tilde{\alpha}} $ and $A=2\pi {r}_h$ is the horizon area of the acoustic black hole. 
{By analyzing, the result for the entropy we find that the second term is a correction of logarithmic type and
the third term shows a correction for the entropy of area that is proportional to the Hawking temperature. 
%$ \tilde{T}_h=\tilde{\alpha}^{3/2}T_h $, of the acoustic black hole. 
Next we display the expression for the entropy in terms of ${r_h}$ given by
\begin{eqnarray}
\label{entropy-bc}
\tilde{S}= {S}{\tilde{\alpha}}=\frac{2\pi {r}_h}{4\sqrt{\tilde{\alpha}}}-\frac{\pi\lambda}{8}\ln{\frac{2\pi {r}_h}{4}}
-\frac{\pi^2\lambda^2}{8}\tilde{\alpha}^{3/2}{T}_h+\frac{\pi\lambda}{8}\ln{[1+2\alpha]}+\cdots.
\end{eqnarray}
For $ \lambda=0 $ and $\tilde{\alpha}=1$ $ (\alpha=0) $ we have, $ S=A/4=2\pi r_h/4 $. 
It should be noted that the last term in the above equation presents a logarithmic correction term that does not depend on the radius of the horizon and shows a dependence on the conserved charge $c=e(1+2\alpha)$.
This charge can be read off from the equation of motion for the gauge field of the Lagrangian (\ref{acao})~\cite{ABP10}, ie }
\begin{eqnarray}
(\nabla \cdot \vec{E})=2e\rho(1+2\alpha)\omega.
\end{eqnarray}

\subsection{Rotating acoustic black hole ($B\neq 0 $)}

\subsubsection{Result with GUP}

{Here we will explore the case with rotation and determine the corrections for the temperature and entropy that arise due to the GUP. In that sense, following all the steps as described above, we find the respective results for temperature and entropy keeping terms up to first order in $ \alpha $:
%\begin{eqnarray}
%{\cal T}_{hGUP}={\cal T}_h\left[ 1-\frac{\lambda}{4\tilde{r}_h }+ \frac{\lambda^2}{8\tilde{r}_h^2 }+\cdots   \right]^{-1}.
%\end{eqnarray}
%Which can also be organized, keeping terms up to first order in $ \alpha $, in the form
\begin{eqnarray}
{\cal T}_{hGUP}&=&\frac{\tilde{\alpha}}{2\pi\tilde{r}_h}
\left[ 1 - T_2
-\lambda\left(\frac{1}{4\tilde{r}_h } - T_3\right)
+\lambda^2\left(\frac{1}{8\tilde{r}^2_h } - \frac{T_3}{2\tilde{r}_h } \right)+\cdots\right]^{-1},
\\
&=&\frac{\tilde{\alpha}}{\tilde{A}}
\left[ 1 - T_2
-\lambda\left(\frac{\pi}{2\tilde{A} } - T_3 \right)
+\lambda^2\left(\frac{\pi^2}{2\tilde{A}^2 } - \frac{\pi T_3}{\tilde{A} } \right) \cdots \right]^{-1},
\end{eqnarray}
where
\begin{eqnarray}
T_2&=&\frac{3\alpha B^2}{2\tilde{r}^2_h }\left[\sqrt{\tilde{\alpha}} +\frac{B}{\tilde{r}_h } \right]
=\frac{\pi^2\alpha B^2}{\tilde{A}^2 }\left[6\sqrt{\tilde{\alpha}} +\frac{2\pi B}{\tilde{A} } \right],
\\
T_3&=&\frac{3\alpha B^2}{8\tilde{r}^3_h }\left[\sqrt{\tilde{\alpha}} +\frac{B}{\tilde{r}_h } \right]
=\frac{3\pi^3\alpha B^2}{\tilde{A}^3 }\left[\sqrt{\tilde{\alpha}} +\frac{\pi B}{2\tilde{A} } \right] .
\end{eqnarray}
For entropy we have
\begin{eqnarray}
\label{entropy-gup}
S_{GUP}&=&\tilde{\alpha}^{-1}\left[\frac{\tilde{A}}{4} + S_1-\lambda\left(\frac{\pi}{8}\ln{\frac{\tilde{A}}{4}}+S_2\right)
-\lambda^2\left(\frac{\pi^2}{32 \tilde{A}/4}+S_3\right)+\cdots\right].
\end{eqnarray}
where
\begin{eqnarray}
S_1&=&\frac{\pi^2\alpha B^2}{16(\tilde{A}/4) }\left[6\sqrt{\tilde{\alpha}} +\frac{2\pi B}{8(\tilde{A}/4) } \right],
\\
S_2&=&\frac{3\pi^3\alpha B^2}{4^3(\tilde{A}/4)^2 }\left[\frac{\sqrt{\tilde{\alpha}}}{2} +\frac{\pi B}{24(\tilde{A}/4) } \right],
\\
S_3&=&\frac{3\pi^3\alpha B^2}{4^4(\tilde{A}/4)^3 }\left[\frac{\sqrt{\tilde{\alpha}}}{3} +\frac{\pi B}{32\tilde{A}/4 } \right] .
\end{eqnarray}
Thus, taking $ B=0 $ reduces to the result of equation (\ref{entA}).
Note that the contributions obtained for the terms, $ S_2 $ and $ S_3 $ due to the GUP have corrections that are only powers of $ 1/\tilde{A} $. 
However, logarithmic corrections of type $\lambda B\ln\tilde{A} $ are not generated.
}

\subsubsection{Result with modified dispersion relation}

{At this point, we shall consider the wave equation in the curved space, i.e., in the background of the metric (\ref{ECC}), to obtain in terms of momentum and energy the following equation~\cite{ABP}
\begin{eqnarray}
\left[1+2\alpha(v_x+v_y)\right]\tilde{{\cal E}}^2+2\left(\vec{v}\cdot\vec{k}\right)\tilde{{\cal E}}-\left(\tilde{\alpha}c^2_s -v^2\right)k^2=0.
\end{eqnarray}
Thus solving the above equation and keeping terms up to first order in $ \alpha $ we find the dispersion relation in the non-relativistic limit to the velocity profile (\ref{vab}) in the vicinity of the event horizon, $ r\rightarrow \tilde{r}_h $, in the form
\begin{eqnarray}
\label{rdmvl}
\tilde{{\cal E}}=E\left(1-2\alpha\sqrt{\tilde{\alpha}} - \frac{2\alpha B}{2\tilde{r}_h}\right) + {\cal O}(\alpha^2),
\end{eqnarray}
where we have considered $ c_s=1 $ and $ E=k $ is the dispersion relation for $ \alpha=0 $.
Next, in order to compare the relationship above with equation (\ref{dispdh}) we can write the dispersion relation (\ref{rdmvl}) as
\begin{eqnarray}
\label{rdmvlq}
\tilde{{\cal E}}^2&=&(1-2\alpha\sqrt{\tilde{\alpha}})^2E^2\left(1 +\frac{\eta\alpha B}{(1-2\alpha\sqrt{\tilde{\alpha}})\tilde{r}_h}\right)^2
\\
&=&(1-4\alpha)E^2\left(1 +\frac{\eta\alpha B}{\tilde{r}_h}\right) +{\cal O}(\alpha^2).
\end{eqnarray}
And then using $ k\sim \Delta k \geq 1/\Delta x=1/\tilde{r}_h $ and redefining $\bar{{\cal E}}^2=\tilde{{\cal E}}^2/(1-4\alpha)  $ we rewrite the above equation as 
\begin{eqnarray}
\label{rdmvlq2}
\bar{{\cal E}}^2= \left(1 +\eta 2\alpha B k\right) E^2+{\cal O}(\alpha^2),
\end{eqnarray}
{which can also be written in terms of the energy difference as}
\begin{eqnarray}
\label{edif}
\frac{\Delta E}{E}=\frac{ \bar{{\cal E}}- E}{E}=\eta \alpha B k.
\end{eqnarray}
Here $ \eta=\pm 1 $ are the polarizations. By comparing with equation (\ref{dispdh}) we identify $ {2\alpha B}k=k^2/\Lambda^2 $, and so we note that the term containing the rotation parameter $ B $ plays the role of a dispersion source associated with the viscosity of the fluid at a momentum scale $k^*=2\alpha B\Lambda^2$. 

In order to emphasize the effect of $ B $ we again consider the Rayleigh's formula (\ref{rf}) 
{and for the dispersion relation (\ref{rdmvlq2}) we find the following velocity difference}
\begin{eqnarray}
\frac{v_g-v_p}{v_p}=\eta\alpha Bk.
\end{eqnarray}
For $ \eta=+1 $ we have the supersonic $ (v_g > v_p)$ case and $ \eta=-1 $ the subsonic $(v_g < v_p)$ case.

Again, applying the Hamilton-Jacobi method as done previously we have
\begin{eqnarray}
\tilde{\cal T}_h={\cal T}_h\left(1 - 2\alpha\sqrt{\tilde{\alpha}} - \frac{2\alpha B}{2\tilde{r}_h}\right)^{-1}
={\cal T}_h\left(1 - 2\alpha\sqrt{\tilde{\alpha}} - \frac{2\pi\alpha B}{\tilde{A}}\right)^{-1}.
\end{eqnarray}
We can also write this equation in the form
\begin{eqnarray}
\bar{\cal T}_h={\cal T}_h\left(1 + \alpha B k\right) +{\cal O}(\alpha^2),
\end{eqnarray}
where we have redefined  $\bar{\cal T}_h={\tilde{\cal T}_h}/{(1+2\alpha)}  $. 
 The above equation can be rewritten in terms of the Hawking temperature variation with respect to changing the medium due to rotation as follows
\begin{eqnarray}
\frac{\Delta {\cal T}_h}{{\cal T}_h}=\frac{\bar{\cal T}_h-{\cal T}_h}{{\cal T}_h}=\frac{\alpha B}{\tilde{r}_h}=\alpha B k.
\end{eqnarray}
{So, we can write the following formula}
\begin{eqnarray}
\frac{v_g-v_p}{v_p}=\frac{\Delta E}{E}={\eta}\frac{\Delta {\cal T}_h}{{\cal T}_h}.
\end{eqnarray}
The formula above may conduct to the study of the Hawking temperature variance in terms of the parameters of the model. For instance in BEC physics \cite{casana,taylor} the deviation of an analogous Lorentz-violating parameter $\beta\sim\alpha B k^*\sim 10^{-5}$ --- see also \cite{Alfaro:2004aa} for related issues in quantum gravity. This leads to an interesting deviation of temperature 
\begin{eqnarray}
\frac{\Delta {\cal T}_h}{{\cal T}_h}\sim 10^{-5},
\end{eqnarray}
which is analogous to the celebrated small temperature fluctuations in the anisotropy of the cosmic microwave background (CMB).

Let us now return to corrected entropy issues. Through the use of the dispersion relation discussed above we obtain the entropy 
\begin{eqnarray}
{\cal S}=\tilde{\alpha}^{-1}\left[\frac{\tilde{A}(1 - 2\alpha\sqrt{\tilde{\alpha}})}{4}-\frac{2\pi\alpha B}{4}\ln{\frac{\tilde{A}}{4}}
+\frac{3\pi^2\alpha B^2 }{\sqrt{\tilde{\alpha}}\tilde{A}}+\frac{6\pi^3\alpha B^3 }{4\tilde{\alpha}\tilde{A}^2}+\cdots\right],
\end{eqnarray}
that can also be presented in terms of the horizon radius as
\begin{eqnarray}
{\cal S}=\tilde{\alpha}^{-1}\left[\frac{2\pi\tilde{r}_h(1 - 2\alpha\sqrt{\tilde{\alpha}})}{4}-\frac{2\pi\alpha B}{4}\ln{\frac{2\pi\tilde{r}_h}{4}}+\frac{3\pi\alpha B^2 }{2\sqrt{\tilde{\alpha}}\tilde{r}_h}+\frac{6\pi\alpha B^3}{16\tilde{\alpha}\tilde{r}_h^2}\cdots\right].
\end{eqnarray}
Moreover, the above equation can be displayed in terms of $ r_h $ in the form
\begin{eqnarray}
\label{entropy-mdr}
\tilde{\cal S}={\cal S}\tilde{\alpha}=\frac{2\pi {r}_h(1 - 2\alpha\sqrt{\tilde{\alpha}})}{4\sqrt{\tilde{\alpha}}}
-\frac{2\pi\alpha B}{4}\ln{\frac{2\pi {r}_h}{4}}+\frac{3\pi\alpha B^2 }{2{r}_h}+\frac{6\pi\alpha B^3}{16 {r}_h^2}
+\frac{2\pi\alpha B}{8}\ln({1+2\alpha})+\cdots.
\end{eqnarray}
Therefore, we show that using the dispersion relation derived from the model itself we have obtained a logarithmic correction term for entropy as well as a term that depends on a conserved charge. Note also that the logarithmic corrections arise only when the parameter $ B $ associated with the circulation is nonzero. On the other hand if $ B = 0 $ these logarithmic corrections are not generated even if $ \alpha \neq 0 $.
{Here we highlight that in obtaining the entropy results in equations (\ref{entropy-bc}) and (\ref{entropy-gup}) we have applied the GUP approach to obtain a modified dispersion relation to compute the quantum corrections for the Hawking temperature and entropy. 
On the other hand, in another approach we have focused on a dispersion relation that is modified but the uncertainty principle preserves the standard Heisenberg form, that is, to obtain the entropy given by equation (\ref{entropy-mdr}) we have considered 
the standard uncertainty principle. % to find a modified dispersion relation. 
Such a modified dispersion relation arises due to the effect of Lorentz symmetry breaking terms that were introduced in the Lagrangian (\ref{acao}) that consequently reflects in the modification of the acoustic metric (\ref{ECC}).
}

\section{conclusions}
\label{conclu}

In summary, by considering the GUP and the tunneling formalism via the Hamilton-Jacobi method, using the WKB approximation, we computed the Hawking temperature and
the entropy associated with the rotating acoustic black hole. We found corrections for temperature and entropy. In particular for entropy we have obtained terms of logarithmic type corrections that appears in the leading order and a contribution related to conserved charge.
Furthermore applying a dispersion relation obtained from the model itself, terms of logarithmic corrections were also generated. Thus for rotating acoustic black holes the modified dispersion relation obtained from a Lorentz-violating background seems to be more effective to find logarithmic corrections than the GUP which are usualy obtained to account the Heisenberg uncertainty principle in black holes and string theory. Such rotating acoustic black holes also provide us with an interesting formula for deviation of temperature, which for BEC systems reveals the interesting result: $\frac{\Delta {\cal T}_h}{{\cal T}_h}\sim 10^{-5}$. This clearly reminds the celebrated temperature deviation in the CMB temperature and investigations in this direction should be further addressed.

\acknowledgments

We would like to thank CNPq/PRONEX and CAPES  for partial financial support.
The work by MAA has been supported by the CNPq
project No. 433980/2018-4. F. A. B. and E. P. acknowledge support
from CNPq (Grants No. 312104/2018-9 and
No. 304852/2017-1).

\end{document}